\documentclass[10pt,preprint2]{aastex}

\slugcomment{Not to appear in Nonlearned J., 45.}

\shorttitle{Seismic and photospheric solar radius}
\shortauthors{Haberreiter et al.}

\begin{document}


\title{Solving the discrepancy between the seismic and photospheric solar radius}


\author{M. Haberreiter and W. Schmutz}
\affil{Physikalisch-Meteorologisches Observatorium Davos, World Radiation Center, Dorfstrasse 33, 7260 Davos, Switzerland}
\email{margit.haberreiter@pmodwrc.ch}
\author{A. G. Kosovichev}
\affil{W. W. Hansen Experimental Physics Laboratory, Stanford University, Stanford, CA 94305-4085, USA}
\email{sasha@quake.stanford.edu}

\begin{abstract}
Two methods are used to observationally determine the solar radius:  
One is the observation of the intensity profile at the limb, the other one uses f-mode frequencies to derive a 'seismic' solar radius which is then corrected to optical depth unity. The two methods are inconsistent and lead to a difference in the solar radius of $\sim$0.3\,Mm. Because of the geometrical extention of the solar photosphere and the increased path lengths of tangential rays the Sun appears to be larger to an observer who measures the extent of the solar disk. Based on radiative transfer calculations we show that this discrepancy can be explained by the difference between the height at disk center where $\tau_{\mathrm{5000}}=1$ ($\tau_{\mathrm{Ross}}=2/3$) and the inflection point of the intensity profile on the limb. We calculate the intensity profile of the limb for the MDI continuum and the continuum at 5000\,{\AA} for two atmosphere structures and compare the position of the inflection points with the radius at $\tau_{\mathrm{5000}}=1$ ($\tau_{\mathrm{Ross}}=2/3$). The calculated difference between the 'seismic' radius and the inflection point is $0.347\pm 0.06$\,Mm with respect to $\tau_{\mathrm{5000}}=1$ and $0.333\pm 0.08$\,Mm with respect to $\tau_{\mathrm{Ross}}=2/3$. We conclude that the standard solar radius in evolutionary models has to be lowered by $0.333\pm 0.08$\,Mm and is 695.66~Mm. Furthermore, this correction reconciles inflection point measurements and the 'seismic' radii within the uncertainty.
\end{abstract}





\section{Introduction}

The solar radius is a key parameter for understanding the
Sun's evolution and many astrophysical applications.
Being our closest star, the Sun can be considered as our
laboratory, allowing us to determine its radius with
high precision. Over the past 25 years various groups were
dedicated to determine the solar radius by applying basically two approaches. First, there is the direct 
measurement of the photospheric radius. In earlier times its visual determination was carried out by an observer. Later, employing CCD-imaging devices, the visual radius determination was replaced by deriving the inflection point to the intensity profile of the limb. The error of both methods has been
investigated in detail by \cite{Laclare1996} who derive a solar radius of 959''.60, which agrees well with 959''.63 $\pm$ 0''.10 or 695.99 $\pm$0.07\,Mm \citep{Allen1976}, commonly used for calibrating
solar evolutionary models \citep{Brown1998,Turck2000}. Second, \cite{Schou1997,Antia1998} determine the helioseismic radius by means of f-mode frequencies obtained from the Michelson Doppler Imager \citep[MDI]{Scherrer1995} on board the ESA/NASA spacecraft SOHO. Due to the high quality of the MDI observations the helioseismic determination is very precise and leads to a radius that is approximately 0.3~Mm smaller than the standard value by \cite{Allen1976}. \cite{Schou1997} conclude that the standard solar radius has to be decreased by this value. Finally, \cite{Brown1998} determine the solar radius from observations obtained with the High Altitude Observatory's Solar Diameter Monitor \citep[HAO SDM]{Brown1982} by applying a model of the data reduction procedure. The authors point out that there are significant uncertainties associated with the currently used radius value which are related to the definition of the solar limb adopted in the radius determinations and the reduction of the measured value to the photosphere. The authors indicate a height of about 0.3\,Mm which 'perhaps' explains the radius correction inferred from the f-mode frequencies. In the present paper we determine the difference between the height at $\tau_{\mathrm{5000}}=1$ at disk center and the distance of the inflection point from radiative transfer calculations for two wavelengths and atmosphere structures. 
\section{Calculations}

With the radiative transfer code {\sc{cosi}} we
calculate the solar limb function for the MDI
continuum intensity and the continuum at 5000\,{\AA}. {\sc{cosi}} is a combination of a model
atmosphere code in spherical symmetry, developed by
\cite{SchmuHamWess1989}
and the spectrum synthesis program {\sc{synspec}}, going
back to \cite{Hubeny1981} and further developed by
\cite{HubenyLanz1992,HubenyLanz1992b}.
The adaption to the solar atmosphere has been published in
\cite{HaberreiterSchmutz2003}. The model atmosphere code calculates the NLTE population numbers for a set of specified atomic levels by solving the radiative transport equations
simultaneously with the equations for statistical
equilibrium. The radiative transfer is solved along rays parallel to the central ray incident on a spherical
distribution of the physical atmosphere structure \citep{Mihalas1975}.
A spherical geometry allows to calculate geometrically correct
the emerging intensity at the limb and line of sights beyond the solar limb.

We base the model calculations on two model atmosphere structures for the quiet Sun, the LTE structure by \cite[hereafter Kurucz91]{Kurucz1991} and the NLTE structure Model C by \citet[hereafter Fontenla07]{Fontenla2007}. 
The radius at disk center where $\tau_{\mathrm{5000}}=1$, is set to the solar radius derived from seismic observations, $R_{\odot}$=695.68~Mm as given by \citet{Schou1997}. The high resolution Ni line profile has been convolved with the five MDI filter functions F0 to F4 used to determine the actual continuum intensity observed by MDI. 
%
\section{Results}
\begin{figure*}
\begin{center}
\epsscale{1.8}
\plotone{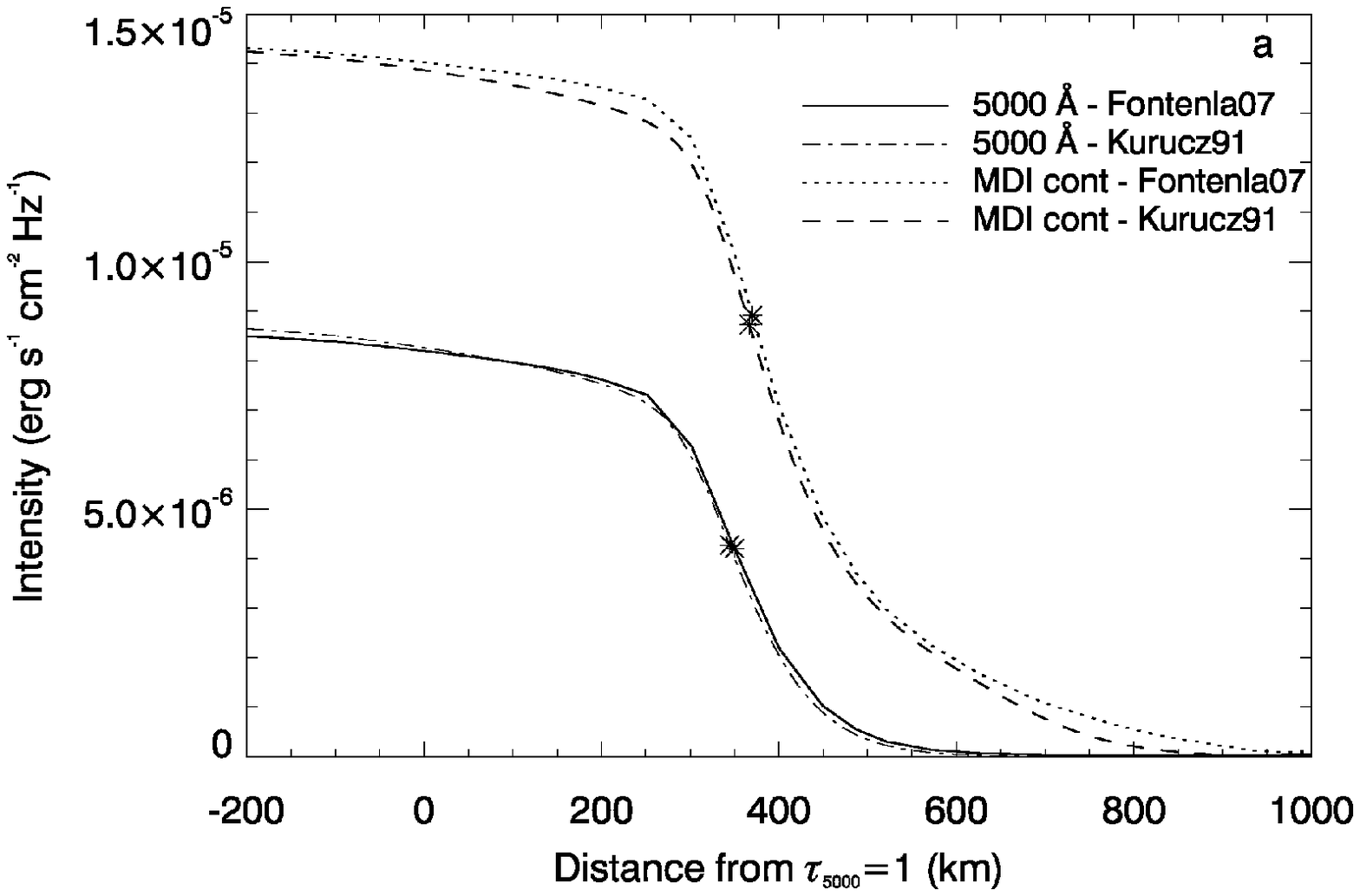}\\
\plotone{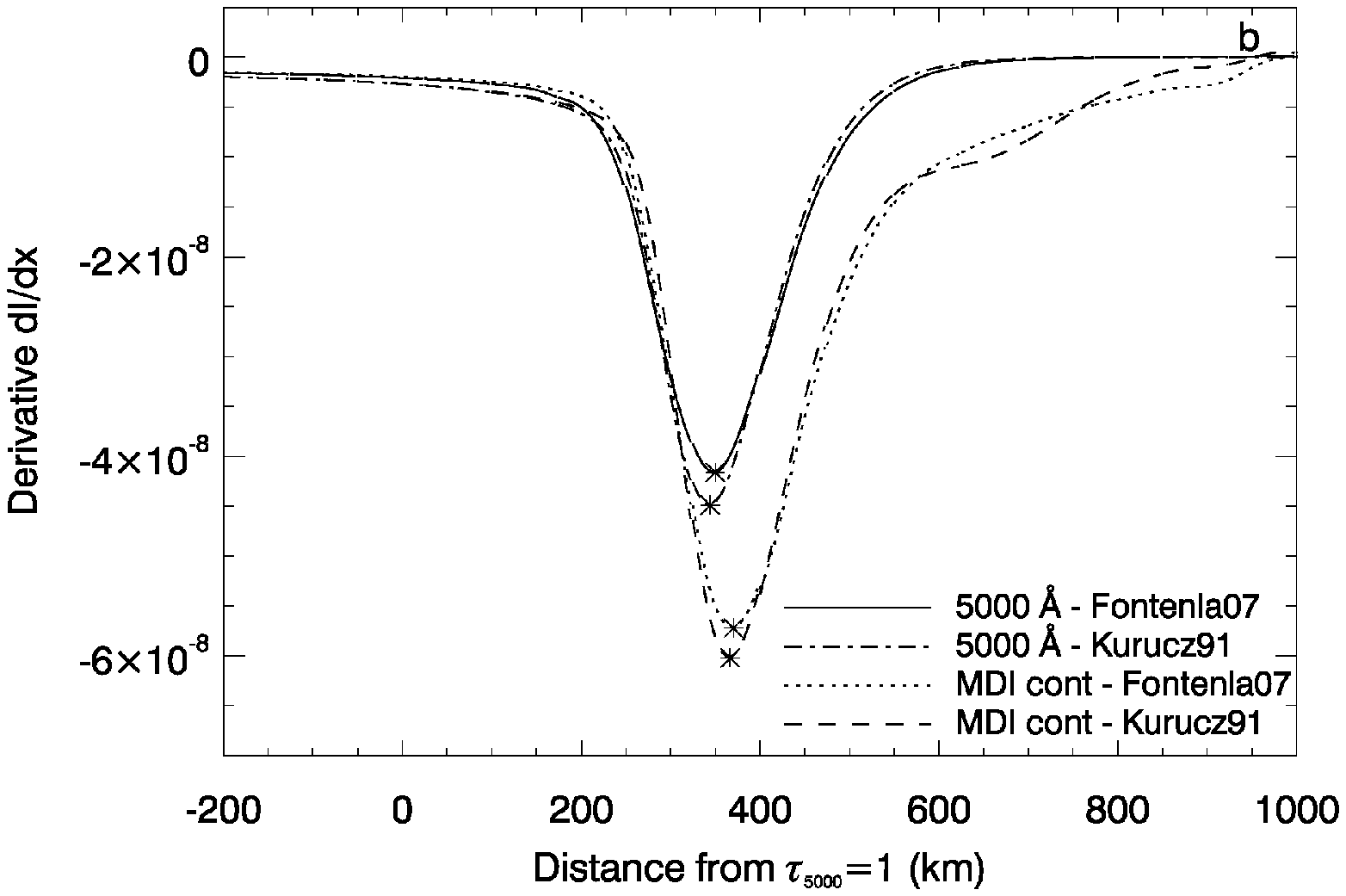}
\caption{Panel (a) gives the intensity profile calculated with the Fontenla07 and Kurucz91 solar atmosphere structures for the 5000\,{\AA} continuum (dotted and dashed-dotted) and the MDI continuum (solid and dashed). Panel (b) shows the derivative of the four intensity profiles. The local minimum is the position of the inflection point, which is 365~km for MDI and 347~km for 5000\,{\AA} (see Table\,\ref{tab:inflex}). Note that the distance to $\tau_{\mathrm{Ross}}=2/3$ is 14~km less.  \label{fig:Limb}}
\end{center}
\end{figure*}

Fig.\,\ref{fig:Limb} panel (a) shows the mean intensity variation for the MDI continuum at the solar limb applying all five MDI filter functions and the continuum at 5000\,{\AA} calculated with the LTE and NLTE model atmosphere structures. Distance zero refers to the 'seismic' solar radius $R_{\odot}$=695.68~Mm, i.e. $\tau_{\mathrm{5000}}=1$ at disk center. 
For the exact determination of the inflection point we calculate the derivative $dI/dx$ ($x$ being the outward coordinate) of all four limb profiles, shown in Fig.\,\ref{fig:Limb} panel (b). The two atmosphere structures agree well in the photosphere and lead to very similar intensity profiles. The derivative of the intensity is always negative and shows a local minimum at the inflection point. The distances of the local minima for all four limb profiles are given in Table\,\ref{tab:inflex}. The mean value of the distance for the two atmosphere structures for $\lambda = 5000$\,\AA\ is 0.347$\pm$0.06~Mm, where we used the difference between the two atmosphere structures as criterion to derive the error introduced by the atmosphere structures. This result is close to the value of 0.370\,Mm estimated by \cite{Wittmann1974} based on the atmosphere structure by \cite{Holweger1967}. 

In evolutionary models the radius is typically defined as the point where the temperature equals the effective temperature, i.e. where $\tau_{\mathrm{Ross}}=2/3$. From our calculations we find that the height where $\tau_{\mathrm{Ross}}=2/3$ is in the mean 0.014~Mm further out than where $\tau_{\mathrm{5000}}=1$. Thus the correction of an inflection point measurement to the 'seismic' radius is 0.333~Mm. This correction explains the discrepancy between the modeled and observed f-mode frequencies. We conclude that the standard radius used in the evolutionary models has to be corrected by value this value, leading to 695.66~Mm.
\begin{table}[ht]
\begin{center}
\caption{Distance $z$ (Mm) between the inflection point and the $\tau_{\mathrm{5000}}=1$ radius ($R_{\odot}$=695.68~Mm) for two different atmosphere structures and wavelengths. The distance to $\tau_{\mathrm{Ross}}=2/3$ is in the mean 0.014~Mm less.\label{tab:inflex}}
\begin{tabular}{cr@{.}lr@{.}lr@{.}lr@{.}l}
\tableline\tableline
  &  \multicolumn{4}{c}{$\tau_{\mathrm{5000}}=1$} \\

Atmosph. struct.  &  \multicolumn{2}{c}{MDI cont.} &  \multicolumn{2}{c}{5000 {\AA} cont.}\\
\tableline

	Kurucz91   & 0&367& 0&344		\\
        Fontenla07 & 0&370& 0&350	   \\

 	\noalign{\smallskip}
\tableline
 	\noalign{\smallskip}
	$\Delta{z}$& 0&003&   0&006    	\\
	$\bar{z}$  & 0&368&   0&347   \\
\tableline
\end{tabular}
\end{center}
\end{table}

\section{Discussions}
\begin{table*}[tt!]
\begin{center}
{\normalsize{
\caption{Solar radii and their mean values $\bar{R}_\odot$ derived by means of i) the standard solar radius and recent inflection point measurements ii) values referring to the 'seismic' radius. As the result given by \cite{Wittmann1997} more than 8$\sigma$ different from the mean we excluded this value from our considerations. \label{tab:radii}}
\begin{tabular}{l r@{.}l@{$\pm$}rlllc}
\tableline\tableline
 	\noalign{\smallskip}	
	Authors  & 
	\multicolumn{3}{c}{R$_\odot$ (Mm)}&
      	R$_\odot$ (arcsec) &
	Description&
	$\bar{R}_\odot$ (Mm)\\
 	\noalign{\smallskip}
	\tableline
 	\noalign{\smallskip}
	\multicolumn{6}{l}{\em{i) IP measurements}}\\
 	\noalign{\smallskip}
	\cite{Allen1976}  & 695&990&0.070 & 959''.63$\pm$0''.02 & standard value  && \\
	\cite{Sofia1994}  & 695&917&0.043 & 959''.53$\pm$0''.06  & SDS && \\
	\cite{Neckel1995} & 695&982&0.022 & 959''.62$\pm$0''.03  & McMath ST&& \\
	\cite{Laclare1996}& 695&830&0.007 & 959''.41$\pm$0''.01  & OCA astrolab&&    \\
	\cite{Kuhn2004}   & 695&740&0.110 & 959''.29$\pm$0''.15  & MDI && \\
	\cite{Emilio2005} & 695&946&0.029 & 959''.57$\pm$0''.04 & SP astrolab &695.901$\pm$0.098 \\
 	\noalign{\smallskip}
 	\noalign{\smallskip}
	\cite{Wittmann1997}& 696&715&0.029 & 960''.63$\pm$0''.04 & Obs. del Teide&&    \\
 	\noalign{\smallskip}
        \tableline
 	\noalign{\smallskip}
	\multicolumn{6}{l}{\em{ii) 'seismic' radii}} \\
	\cite{Schou1997} & 695&680&0.030 & 959''.20$\pm$0''.01& MDI f-modes& \\
	\cite{Antia1998} & 695&787& - & 959''.35$\pm$ - &GONG f-modes&   \\
	Brown \& CD (1998) & 695&508&0.026 & 958''.97$\pm$0''.04& HAO SDM &695.658$\pm$0.140			 \\
 	\noalign{\smallskip}
\tableline
\end{tabular}
}}
\end{center}
\end{table*}
\begin{table}[t!]
\begin{center}
{\normalsize{
\caption{Correction of 0.333~Mm applied to the mean of the inflection point measurements. The corrected radius value is compared with the mean of the 'seismic' radii given in Table \,\ref{tab:radii}, referring to $\tau_{\mathrm{Ross}}=2/3$. The result is consistent within the uncertainty. \label{tab:radii2}}
\begin{tabular}{lr}
\tableline\tableline
 	\noalign{\smallskip}	
	& 
 	Radius (Mm)\\
 	\noalign{\smallskip}
	\tableline
 	\noalign{\smallskip}
	$\bar{R}_{\mathrm{IP}}$ 	&695.901$\pm$0.098   \\
	Correction			&  0.333$\pm$0.008   \\
	Corrected $\bar{R}_\odot$ 	&695.568$\pm$0.098   \\
	$\bar{R}_\odot$  		&695.658$\pm$0.140   \\
	$\Delta R$  		&          0.090$\pm$0.171  \\
 	\noalign{\smallskip}
\tableline
\end{tabular}
}}
\end{center}
\end{table}
For consistency we investigate whether recent inflection point measurements are compatible with the mean of 'seismic' radii relating to $\tau_{\mathrm{Ross}}=2/3$, i.e. where the temperature equals the effective temperature, by applying this correction term. Table \ref{tab:radii} gives the solar radii based on ii) IP measurements, ii) 'seismic' radii. The mean values deviate by $\sim$0.24~Mm. In Table\,\ref{tab:radii2} we correct the mean radius values of the inflection point observation by 0.333~Mm and compare the result with the mean 'seismic' radius, leading to a difference of 0.09$\pm$0.17~Mm. We conclude that the inflection point measurements and the 'seismic' radius can be reconciled by applying the correction of 0.333~Mm.

Our calculations indicate that the position of MDI continuum inflection point is by 0.02~Mm further out than for 5000\,{\AA}. The radius value determined by \cite{Kuhn2004} (see Table\,\ref{tab:radii}) refers to the inflection point of the intensity profile observed by MDI. The difference of 0.02~Mm cannot explain the somewhat lower radius value by \cite{Kuhn2004}. 

We emphasis that we do not consider any variation of the solar diameter with time \citep{Emilio2007,ThuiHabSab2005,Kuhn2004,Toulmonde1997,Parkinson1980}, which however could explain some of the deviation of the radius values. Furthermore, our considerations are with respect to the quiet Sun. 
\section{Conclusions}
From radiative transfer calculations with the spherical code COSI we determine a distance of 0.333$\pm$0.008~Mm ($0.347\pm 0.06$\,Mm) between the height where $\tau_{\mathrm{Ross}}=2/3$ ($\tau_{\mathrm{5000}}=1$) at disk center and the position of the inflection point of the 5000\,{\AA} intensity profile at the limb. This correction explains the differences between the f-mode frequencies derived from models calculations and observations. Therefore, we conclude that standard solar radius currently used in evolutionary models has to be corrected by 0.333~Mm and is 695.66~Mm.

Furthermore, the correction of 0.333~Mm reconciles the radius values derived from inflection point measurements and the 'seismic' radius by 0.09$\pm$0.17~Mm. Due to the increasing precision of the instruments it becomes necessary that the community agrees on a generally binding definition of the solar radius and that values derived via different techniques are to be corrected accordingly. We propose that the solar radius refers to $R_\odot=R(\tau_{\mathrm{Ross}}=2/3)$. Thus, we suggest that the value of the solar radius determined from inflexion point measurements has to be lowered by 0.333~Mm to be comparable with 'seismic' radii. A detailed study of the position of the wavelength and solar cycle dependent inflection point will be crucial for upcoming observations with the SODISM instrument on board PICARD \citep{ThuiDewSchm2006ASpR}.
\begin{acknowledgements}
This work has been supported by the SNF with the Project No. 200020-109420.
\end{acknowledgements}

\end{document}